\documentclass[prl,aps,final,showpacs,twocolumn,floatfix]{revtex4}
\input pstricks \input epsf
\begin{document}
\def\beq{\begin{equation}}
\def\eeq{\end{equation}}

\title{Kinetic approach to the cluster liquid-gas transition}

\author{F. Calvo}
\affiliation{Laboratoire de Physique Quantique, IRSAMC, Universit\'e Paul
Sabatier, 118 Route de Narbonne, F31062 Toulouse Cedex, France}

\begin{abstract}
The liquid-gas transition in free atomic clusters is investigated
theoretically based on simple unimolecular rate theories and assuming
sequential evaporations. A kinetic Monte Carlo scheme is used to
compute the time-dependent properties of clusters undergoing multiple
dissociations, and two possible definitions of the boiling point are
proposed, relying on the cluster or gas temperature. This numerical
approach is supported by molecular dynamics simulations of clusters
made of sodium atoms or C$_{60}$ molecules, as well as simplified rate
equation.
\end{abstract}
\maketitle

Boiling is often considered as archetypal of first-order phase
transitions in bulk matter. Recent experimental evidence
\cite{hablgt,farizon,brechsr,chabot} suggests that the liquid-gas
transition also occurs in finite atomic systems. The
caloric curves measured for sodium \cite{hablgt}, hydrogen
\cite{farizon} or strontium \cite{brechsr} clusters exhibit a plateau
or a backbending, believed to be signatures of a phase
transition rounded by size effects. Similar conclusions have been
inferred from collisions between gold nuclei \cite{pochodzalla95}. The
experiments performed by Pochodzalla and coworkers have since
motivated a significant amount of theoretical work to help the search
for a possible equation of state for nuclear matter
\cite{gross,belkacem,campi,dorso,chomaz}.

For more than two decades, fragmentation has been recognized as one of
the most convenient ways of accessing fundamental cluster properties
such as binding energies or temperatures. These quantities were
related to each other through calorimetric experiments on the
solid-liquid phase change \cite{habslt,jarrold}. The crucial role of
the observation time was also noticed at an early stage, and more fully
understood
by Klots \cite{klots} who introduced the concept of the evaporative
ensemble. Free clusters are never strictly stable when their energy
exceeds a certain dissociation threshold, as fragmentation will occur,
possibly very late. Measurements on free clusters are thus conducted
on species resulting from hotter and bigger clusters, and which
have sufficiently cooled down so as not to evaporate further. Unfortunately,
the relatively long times involved in experiments (usually the
$\mu$s--ms range) have prevented Klots' ideas from being exploited in later
theoretical studies. While melting can be conveniently addressed
using Monte Carlo (MC) or molecular dynamics (MD) simulations by
artificially keeping the cluster in a container \cite{lba},
fragmentation of a cluster into vacuum is an out-of-equilibrium
phenomenon. The difficulty of accounting for the time variable
has generally been circumvented by considering lattice-gas or
percolation models \cite{campi,chomaz} as well as periodic
boundary conditions \cite{gross,dorso}.

Unimolecular rate theories provide a general framework to describe
single dissociation events accurately and over long time
scales \cite{wa}. Here we introduce a kinetic Monte Carlo
(KMC) scheme based on such rate theories to calculate the caloric
curves of clusters across the liquid-gas transition for arbitrarily
long observation times.

\begin{figure}[htb]
\setlength{\epsfxsize}{8cm} 
\leavevmode\centerline{\epsffile{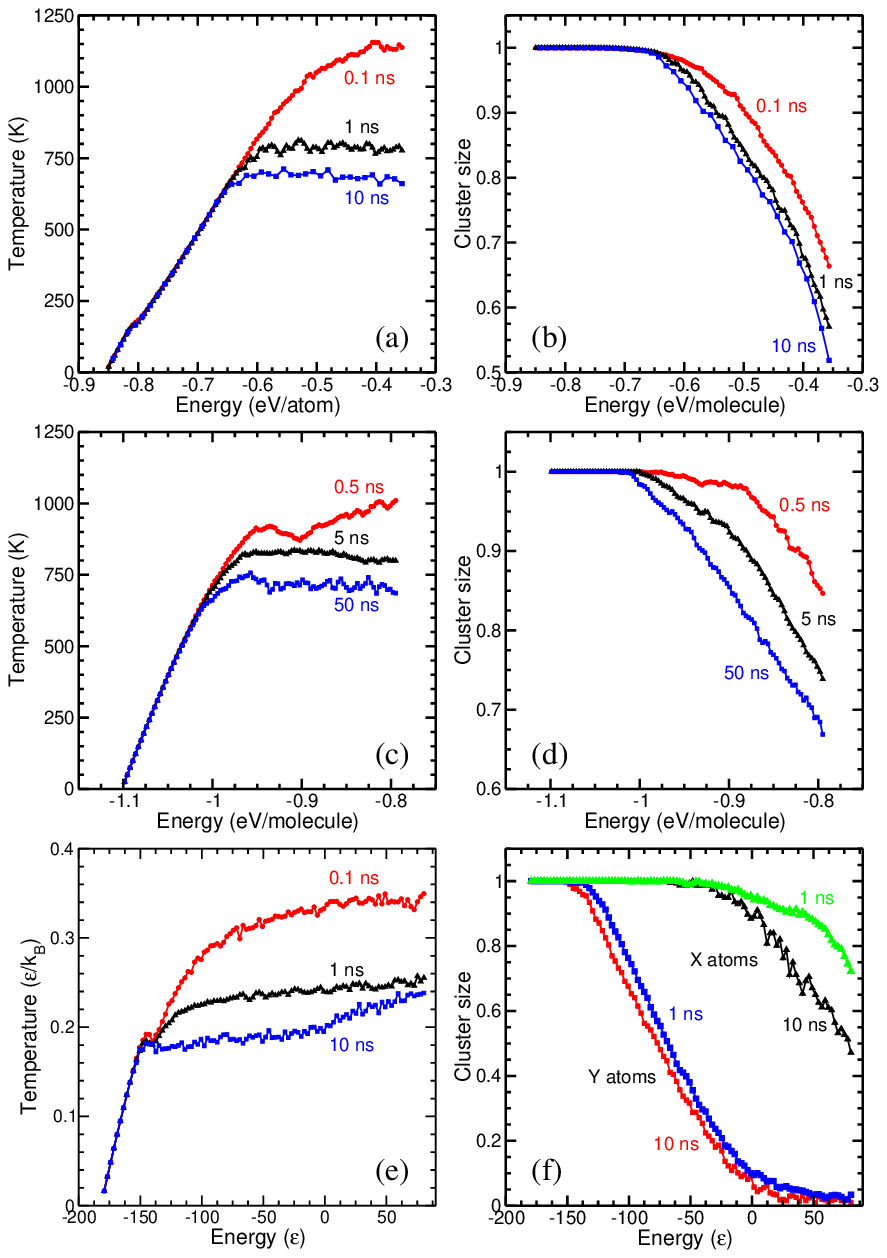}}
\caption{Caloric curves and average cluster size after various waiting
times in MD simulations. (a) and (b) Na$_{55}$; (c) and (d)
(C$_{60}$)$_{40}$; (e) and (f) binary LJ cluster X$_{13}$Y$_{42}$.
Cluster sizes are given relative to the initial size.}
\label{fig:md}
\end{figure}

We start by illustrating the boiling problem by showing the results of
classical MD simulations carried out on three selected clusters,
namely Na$_{55}$, (C$_{60}$)$_{40}$, and a model binary cluster
X$_{13}$Y$_{42}$. These systems are described using an explicit
many-body potential \cite{napot}, the Girifalco potential
\cite{girifalco}, and
Lennard-Jones (LJ) interactions, respectively. For the LJ cluster,
we chose $\varepsilon_{\rm XX}= 1$, $\varepsilon_{\rm YY}=1/2$ and
$\varepsilon_{\rm XY} = 1/\sqrt{2}$, all distance units and masses
being set to one.
For each cluster, MD trajectories are  performed starting from
the lowest-energy minimum, at increasing total energies and zero
angular and linear momenta. After some observation time $t$ we
determine the biggest remaining cluster as the largest set of
connected atoms or molecules, two atoms being connected when their
distance is smaller than a cut-off value $r_{\rm cut}$ chosen as
twice the equilibrium distance. The instantaneous temperature of
this cluster is then calculated after removing the overall translation
and rotation contributions. The average cluster size and temperature
obtained over 100 independent trajectories (20 for the longest waiting
times) are represented for the three clusters in Fig.~\ref{fig:md}.

All caloric curves exhibit two clearly distinct regimes. At low
energies, the roughly linear increase of temperature characterizes
the condensed phases as solidlike and liquidlike. For the sodium and,
to a lesser extent, for the LJ cluster, a small inflection marks the
onset of melting. At high energies, the cluster temperature drops and
reaches a plateau. The plateau temperature is lower and the change
between the two regimes is sharper as the waiting time increases.
This behavior is consistent with the observed smaller average cluster
size, and results from a stronger evaporative cooling. In the cluster
of fullerene molecules, no evidence for the melting transition is seen
on the caloric curves, in agreement with the lack of stability of the
liquid phase known in bulk C$_{60}$ \cite{fullerite}. The backbending
seen for the shortest waiting time (0.5~ns) was confirmed by another
set of 100 independent MD trajectories. However, without any container
preventing dissociation, it is hard to attribute this feature to
melting or to boiling. The binary LJ cluster was constructed as a two-layer
icosahedron, with the most strongly bound atoms X in the center.
Provided that the waiting time is long enough, X atoms may dissociate,
but only after the Y atoms have evaporated. The fragmentation of
X atoms is shown on the caloric curve corresponding to $t=10$~ns by
the slight increase at high excitation energies.

The time-dependence of fragmentation caloric curves can also be
calculated from simple cluster models, similar to those introduced
by Bixon and Jortner for the isomerization problem \cite{bixon}. Our first
assumption is that clusters are heated adiabatically, allowing
fragmentation to occur through sequential loss of monomers
\cite{dimer}. We use simple rate theories to describe each
dissociation step. The dissociation rate $k_n(E_n)$ of cluster X$_n$ into
X$_{n-1}$ depends on the dissociation energy $\Delta_n$ and the total
energy $E_n$ through the harmonic RRK approach \cite{rrk}, namely
$k_n(E_n)=\nu_0(1-\Delta_n/E_n)^{3n-6}$. This well known formula contains a
single time scale parameter $\nu_0$, related to the typical
vibrational period. Once evaporation has occured, the cluster loses a part of 
its internal energy, which is described more accurately using the
Weisskopf theory \cite{weisskopf}. The probability distribution of
the kinetic energy released (KER), $p_n(\varepsilon,E_n)$, is
$p_n(\varepsilon,E_n)\propto \varepsilon (E_n-\Delta_n -
\varepsilon)^{3n-7}$. The dissociation of
rigid molecules would be described similarly, changing the factor
$\varepsilon$ and the number $3n-7$ of degrees of freedom to
$\varepsilon^{3/2}$ and $5n-7$ for linear molecules, and to
$\varepsilon^2$ and $6n-7$ for tridimensional molecules, respectively.

The problem can be further simplified by assuming that the lifetime
of the parent cluster is $1/k_n$ and that evaporative cooling removes
the average KER $\langle\varepsilon\rangle_n = 2(E-\Delta_n)/(3n-7)$,
leading to the energy of the product X$_{n-1}$, $E_{n-1} = E_n - \Delta_n
- \langle\varepsilon\rangle_n$. The multifragmentation problem is thus
reduced to computing all successive dissociation rates, from which
the survival probabilities $p_n(t)$ of all cluster sizes are
calculated following a master rate equation $dp_n/dt= k_n p_n -
k_{n-1} p_{n-1}$, solved exactly.
A similar ``mean-field'' technique was used by Hervieux {\em et al.} to
calculate branching ratios in the collision-induced fragmentation of
Na$_9^+$ \cite{hervieux}. 

Beyond this approximate treatment, the kinetic Monte Carlo method
\cite{kmc} accounts for the continuous character of dissociation times
and KER distributions. Starting with the parent cluster size $n$ at
total energy $E_n$, the evaporation rate $k_n(E_n)$ is calculated and
the KER $\varepsilon$ is chosen randomly from the
distribution $p_n(\varepsilon,E_n)$. The energy is decreased by
$\Delta_n+\varepsilon$, and the time is updated by the
quantity $-\log({\tt ran})/k_n$, where $0\leq
{\tt ran}< 1$ is a random real number. This process is iterated until 
either the waiting time has been exceeded or the remaining energy is
below the next dissociation limit. The final cluster
has $k$ atoms and the energy $E_k$, its temperature is obtained
in the harmonic approximation $k_B T_k=E_k/(3k-6)$.

\begin{figure}[htb]
\setlength{\epsfxsize}{8cm} 
\leavevmode\centerline{\epsffile{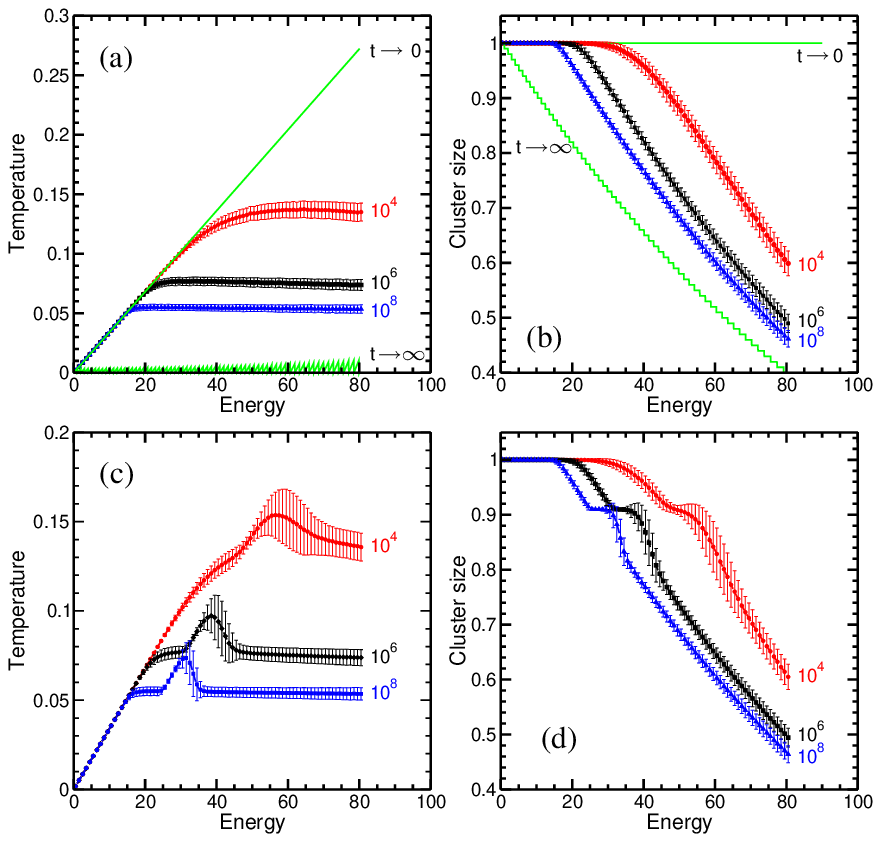}}
\caption{Caloric curves and average final cluster size obtained from
KMC calculations for model clusters without (a and b) or with (c and
d) a specially stable fragment at $n=90$. The waiting times are given
in units of $1/\nu_0$, and the cluster sizes are relative to the
initial size 100.}
\label{fig:model}
\end{figure}

The KMC and mean-field methods have been compared on the simplest case
of the multiple dissociation of a 100-particle model cluster with
$\Delta_n=1$ for all $n$. Figs.~\ref{fig:model}(a) and (b) show
the caloric curve and the average final cluster size obtained from the
KMC calculations. The mean field values were not reported, as they are
undistinguishable from the stochastic data. It should be noted that
these results are not significantly affected when the energetics of
dissociation are described using RRK theory, $p_n(\varepsilon,E_n)
\propto (E_n-\Delta_n - \varepsilon)^{3n-6}$.

The MD results of Fig.~\ref{fig:md} are well
reproduced by the predictions of this simple model, especially
the plateau of the vapor phase and the increasing sharpness of
the transition for long observation times.
The absence of any feature on the caloric curve for $t\to
0$ demonstrates the kinetic character of the cluster liquid-gas
transition. On the other hand the asymptotic curve for $t\to\infty$
exhibits multiple tiny backbendings corresponding to consecutive
dissociations leaving the cluster perfectly cold. We have simulated
the effect of a more stable cluster at size $n^*=90$ by imposing
$\Delta_{n^*}=1.5$. The influence of this magic cluster on the
curves of Figs.~\ref{fig:model}(c) and (d) is rather local. The greater
stability of the X$_{n^*}$ cluster induces a delay on the subsequent
dissociations, but the overall plateau temperatures do not
vary much.

The kinetic Monte Carlo method can be used in more general situations,
for which the rate equation approach is not practical. Clusters with
competing fragmentation channels turn into products having different
internal energies depending on their history. For instance,
heterogeneous clusters X$_n$Y$_p$ can dissociate in multiple ways:
\begin{center}
X$_n$Y$_p$ \\
$\swarrow~\searrow$ \\
X$_{n-1}$Y$_p$ ~~ X$_n$Y$_{p-1}$ \\
$\swarrow~\searrow$ ~~~ $\swarrow~\searrow$ \\
$\cdots$ ~~ X$_{n-1}$Y$_{p-1}$ ~~ $\cdots$
\end{center}
Statistically the X$_{n-1}$Y$_{p-1}$ fragment does not have the same
energy depending on whether it was produced from
X$_{n-1}$Y$_p$ or from X$_n$Y$_{p-1}$. For the above example, and
within the Weisskopf theory, do the two
average energies of the X$_{n-1}$Y$_{p-1}$ product differ by
\begin{eqnarray}
\Delta E_{n-1,p-1} &=& 2(\Delta_{\rm X}-\Delta_{\rm Y}) \nonumber \\
&\times& \displaystyle \frac{3(n+p)-12}{[3(n+p)-7]\cdot [3(n+p)-10]},
\label{eq:de}
\end{eqnarray} 
depending on which of X or Y was emitted first. Only at large sizes
or for similar binding energies $\Delta_{\rm X}\simeq\Delta_{\rm
Y}$ the two fragmentation routes become equivalent. 
Moreover, clusters may be prepared at fixed temperature rather than
fixed energy. The initial energy is then randomly picked
from the canonical distribution $\rho_n(E) \propto E^{3n-6}\exp(-E/k_BT)$
as the first Monte Carlo step.

\begin{figure}[htb]
\setlength{\epsfxsize}{8cm} 
\leavevmode\centerline{\epsffile{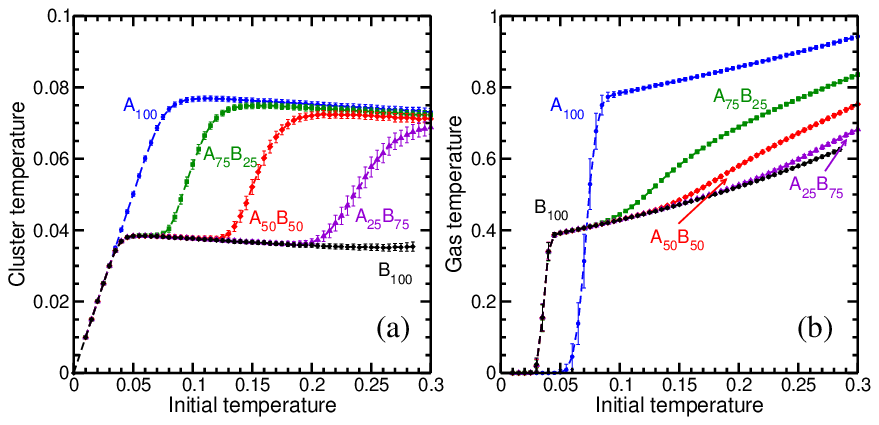}}
\caption{Final cluster (a) and gas (b) temperatures versus initial
temperature in the boiling transition of model binary
clusters with various compositions, for the waiting time $10^6/\nu_0$.}
\label{fig:binary}
\end{figure}

By mimicking the actual dissociation cascade followed by each
individual cluster, the KMC technique offers a realistic statistical
description of multifragmentation. The caloric curves of model binary
clusters X$_n$Y$_{100-n}$ were calculated assuming that X and Y atoms
are bound to the clusters by the energies $\Delta_{\rm X} = 1$ and
$\Delta_{\rm Y}=1/2$, respectively. The caloric curves represented in
Fig.~\ref{fig:binary}(a) for several compositions exhibit two
steps, each associated with the boiling of Y, then X particles. This
two-step liquid-gas transition is somewhat similar to the multi-step
melting process often seen in cluster simulations \cite{calspi}.
As more Y atoms are added, a larger energy is required to
evaporate them, therefore the gas transition of the remaining X
cluster is initiated at higher temperatures. The different onsets
of boiling for X and Y particles in Fig.~\ref{fig:md}(c) and (d)
confirm this two-step transition.

We now address possible definitions of the boiling temperature, based
on the previous results. The onset of the drop in the slope of the
caloric curve provides one possible estimate. However, especially at
short waiting times, the variations of the average remaining cluster
size seem to be more reliable.
Another observable consists of looking at the kinetic energy of the
evaporated atoms, in a fashion somewhat more closely related to
experimental conditions. Using the same notation as above for the initial
and final cluster sizes and energies, and assuming that all
$n-k$  evaporated atoms behave like a perfect gas, a gas temperature
can be defined by $3(n-k)k_BT/2 = E_n-E_k$.

The variations of the
average gas temperature across the boiling transition are shown in
Fig.~\ref{fig:binary}(d) for the model binary clusters. Boiling
is made evident by the sudden increase of the gas temperature, which
provides us with another characterization of the liquid-gas point.
The second boiling point involving the more strongly bound X atoms
does not lead to significant variations of the gas temperature,
because X and Y atoms contribute equally once in the gas phase.
A better estimate of the liquid-gas transition temperature of the
cluster of X atoms is found with the variations of its final size
[see Fig.~\ref{fig:md}(d)].

\begin{figure}[htb]
\setlength{\epsfxsize}{8cm} 
\leavevmode\centerline{\epsffile{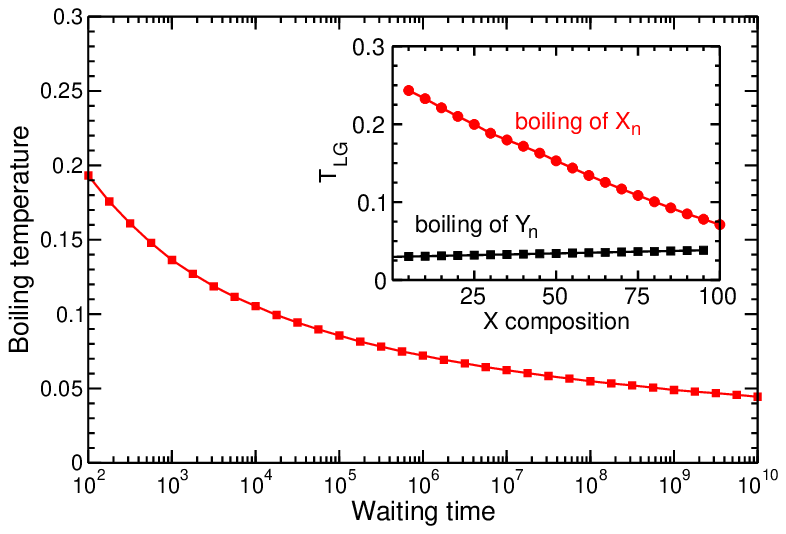}}
\caption{Boiling temperature of model 100-atom homogeneous
cluster versus waiting time. Inset: boiling temperatures of the X and
Y parts of model X$_n$Y$_{100-n}$ clusters versus composition $n$, for
the waiting time $t=10^6/\nu_0$.}
\label{fig:tlg}
\end{figure}

The three aforementioned definitions of the boiling point yield
similar values in a broad range of situations covered by our simple
models. The dependence of the observation time on the boiling
temperature of the homogeneous cluster is represented in
Fig.~\ref{fig:tlg}. As expected, waiting longer favors evaporation,
hence a lower boiling temperature. In most experiments
\cite{hablgt,farizon,brechsr,chabot} the waiting time is
at least 5 orders of magnitude larger than the typical vibrational
period. Fig.~\ref{fig:tlg} suggests that the results of these
experiments should remain stable by less than 5\% if the waiting time
is doubled or halved. However, similar measurements on trapped
clusters might show some deviation since the trapping time may
exceed seconds \cite{parks}.

In Fig.~\ref{fig:tlg} we also show how the presence of more weakly
bound (Y) atoms influence the boiling point of an (X) cluster. The
fragmentation temperature of Y atoms does not significantly change
with their initial composition. However, as more Y atoms are added,
the X product gets progressively colder due to the more numerous
evaporations. This explains why the boiling temperature of the X
cluster decreases with its composition.

The KMC technique outlined in this paper could be improved using more
accurate unimolecular rate theories, such as Phase Space Theory, which
incorporates anharmonic densities of states as well as a rigorous
treatment of angular momentum constraints \cite{wa}.
The main assumption of the present approach is that boiling in slowly
heated clusters occurs sequentially. Our simulation results seem to
validate this hypothesis. As in glasses \cite{kmcglass}, the
arbitrarily long time scales reached by the present statistical
approach make it a useful alternative to molecular dynamics. We
believe that it forms a bridge between unimolecular rate descriptions
and multifragmentation models that assume thermal equilibrium.

As a first application of the KMC method to the cluster
dissociation problem, we have considered the liquid-gas transition.
Our results emphasize the important role played by the observation
time on the caloric curves. They also indicate
that multiple-step boiling transitions could be detected in
heterogeneous clusters. Beyond these examples, a more complete
interpretation of recent experiments \cite{hablgt,farizon,brechsr},
especially on clusters exhibiting competing dissociation channels
\cite{chabot}, could be anticipated.

\end{document}